# Omnidirectional and broadband absorption enhancement from trapezoidal Mie resonators in semiconductor metasurfaces

*Ragip A. Pala[1,2], Serkan Butun[3], Koray Aydin[3], Harry A. Atwater[1,2]*

[1]Thomas J. Watson Laboratories of Applied Physics, California Institute of Technology, United States
[2]Kavli Nanoscience Institute, California Institute of Technology, United States
[3]Department of Electrical Engineering and Computer Science, Northwestern University, United States

**Abstract**

Light trapping in planar ultrathin-film solar cells is limited due to a small number of optical modes available in the thin-film slab. A nanostructured thin-film design could surpass this limit by providing broadband increase in the local density of states in a subwavelength volume and maintaining efficient coupling of light. Here we report a broadband metasurface design, enabling efficient and broadband absorption enhancement by direct coupling of incoming light to resonant modes of subwavelength-scale Mie nanoresonators defined in the thin-film active layer. Absorption was investigated both theoretically and experimentally in prototypes consisting of lithographically patterned, two-dimensional periodic arrays of silicon nanoresonators on silica substrates. A crossed trapezoid resonator shape of rectangular cross section is used to excite broadband Mie resonances across the visible and near-IR spectra. Our numerical simulations, optical absorption measurements and photocurrent spectral response measurements demonstrate that crossed trapezoidal Mie resonant structures enable angle-insensitive, broadband absorption. A short circuit current density of 12.0 mA/cm$^2$ is achieved in 210 nm thick patterned Si films, yielding a 4-fold increase compared to planar films of the same thickness. It is suggested that silicon metasurfaces with Mie resonator arrays can provide useful insights to guide future ultrathin-film solar cell designs incorporating nanostructured thin active layers.

**Introduction**

Optically engineered nanostructures have opened new design paths for solar cells that feature light management as an integral component of cell design[1-6], leading to higher open-circuit voltages, and higher short-circuit currents. In recent years it has also been shown that the ray optical light trapping limit[7] can in principle even be surpassed using ultrathin film cell designs that operate in the wave optics regime.[8-11] Efficient light absorption in ultrathin film solar cells requires both an increase in the number of optical states in the absorber layer across the solar spectrum and an optimal broadband light-coupling scheme. Numerous solar cell designs have been proposed to achieve light management by employing, e.g., plasmonic design[12], photonic crystal architecture[13, 14] and excitation of dielectric waveguide modes[8] and Mie resonances[15], all of which serve to increase the local density of optical states in the active absorber layer. A second important requirement for efficient light trapping is effective coupling into these modes, which is often realized over only a small bandwidth for resonant structures, as compared to the solar spectrum. Separate efforts have been made to develop optical coupler designs for broadband light coupling using non-periodic[16] and disordered[17, 18] structures. However achieving an increase in the bandwidth of the coupling via of disordered systems results in a significant decrease in the coupling efficiency. It is thus worth developing novel design architectures that enable both a high density of optical states and broadband light coupling into ultrathin-film solar cells.

Ultrathin optical materials structured on the subwavelength scale offer an unprecedented opportunity to control the effective optical materials properties[19, 20], phase[21] and amplitude and potentially realize effective light trapping. Nonetheless these structures are composed of resonators with a particular design frequency[22], limiting their performance to narrowband applications. Metasurfaces with multiple resonant spectra can be obtained by interlacing semiconductor nanostructures with distinct resonance frequencies[23], however only a limited number of resonances can be excited across the solar spectrum. In this work, we report a broadband metasurface design, composed of subwavelength, multi-resonant Mie resonators that enhances light trapping and increases absorption with a broadband spectral response. Efficient coupling is achieved to resonant modes of subwavelength-scale nanoresonators incorporated into a thin film crystalline silicon absorber layer. In a Mie resonator, the resonant wavelength and the bandwidth are simply determined by the design parameters of the nanostructure. In our prototype structures, we break symmetry in a subwavelength volume by use of crossed trapezoids[24] with rectangular cross section as absorbers, which enables formation and excitation of multiple Mie and guided mode resonances and achieves broadband light trapping across the visible spectrum.

## Results

**Light Trapping in Resonant Si nanostructures**

To motivate the absorption enhancement mechanism in trapezoidal Mie resonator structures, we focus first on a simple model system consisting of a structured Si thin film on an $SiO_2$ substrate patterned with a periodic array of Si ridges with rectangular (Fig. 1a – top) and trapezoidal shape (Fig. 1a - bottom). Such a high-index dielectric ridge structure act as an array of resonators, with increased absorption arising from two distinct origins: 1) Mie resonances that localize light intensity in individual high-index ridge regions, with the resonant frequencies determined by the size of the trapezoid or wire cross section, and 2) delocalized resonant waveguide modes supported by the entire periodic ridge array in the Si film, with incoupled frequencies determined by the periodic array properties. To investigate these two coupling regimes, we performed full-field simulations and calculated the Si slab absorption for normally incident plane waves. In our simulations, we considered Si ridges of width $w = 85$ nm, thickness $t = 100$ nm, period $P = 400$ nm and a Si thickness of $t_{Si} = 210$ nm. Figure 1b shows the magnetic field profile for a transverse magnetic (TM) polarized plane wave at free space wavelengths of 575 nm and 620 nm respectively. At 575 nm, light is localized within the Si ridges by coupling into Mie resonances. On the other hand, at 620 nm Si ridges enable coupling into the waveguide modes of the Si slab. In both cases the local H-field amplitudes are enhanced by more than an order of magnitude compared to incoming field. To analyze the contributions of these resonances to absorption across the spectrum, we simulate the absorption spectrum of the nanostructured Si film and the bare Si film of an equivalent thickness, $t = 135$ nm (Fig. 1c). The absorption for the bare Si film is highest at short wavelengths and rapidly decreases towards the Si bandgap. The peaks seen for the bare film arise from Fabry-Perot resonances in the Si layer. Several absorption enhancement features can be identified in the absorption spectrum of the stripe array (in green), corresponding to different orders of the Mie and guided mode resonances, yielding in total a 50% overall increase in photocurrent density. At each resonance, the Si ridges act like nanoantennas, altering the flow of the light into the semiconductor, and inducing an increase in light absorption. These resonances can be also identified by analyzing the abrupt phase shifts of the reflected wave from the Si metasurface (Fig. 1d). Despite a significant overall photocurrent enhancement from the rectangular stripe array, we observe that these resonances are quite sparse across the solar spectra.

**Broadband Absorption Mechanism in Trapezoidal Structures**

Broadband light absorption is achieved by using ridges with a trapezoid pattern (plotted in red). Varying the resonator width along the ridges from 20 nm to 120 nm facilitates Mie resonance excitation of the same resonance order at multiple wavelengths. Note that the periodicity of the resonators along the ridges is smaller than the incident beam wavelength which ensures efficient interaction of the Mie resonators with the incoming light, yielding a broadband absorption enhancement. The phase plot of reflection from the Si trapezoid metasurface reveals a broadband suppression of reflection and excitation of multiple resonances with broadband absorption enhancement across the solar spectrum. (Fig. 1d).

To evaluate the proposed design experimentally, we fabricated two-dimensional periodic arrays of Si ridges with trapezoid and rectangular patterns embedded in 210 nm thick Si-on-insulator (SOI) films. A crossed nanostructure array was used to provide polarization independent response. Scanning electron microscopy images of the fabricated arrays are shown in Fig. 1f. The design parameters are chosen to maximize integrated absorption for AM 1.5G illumination, determined using full-field electromagnetic simulations.

For optimum absorption, a periodicity of 450 nm is used for rectangular and trapezoid arrays. The width of the fabricated rectangular stripes are 115 nm while the width of the trapezoid pattern varies from 40 nm to 220 nm; the resonator thickness is 125 nm. To verify our predictions, we performed spectral reflectivity and spectral response photocurrent measurements and full-field simulations on planar and the nanoresonator-patterned silicon-on-insulator films of the same thickness, $t_{Si}$ = 210 nm, without antireflection coatings (Fig. 2).

The measured optical absorption spectra show excellent agreement with electromagnetic simulations (Fig. 2a-c). The high reflection loss from the planar Si film, is partially mitigated by the crossed rectangular resonators whereas over 90% absorption is achieved by the resonators with trapezoid shape at short wavelengths. We note that in this case, the optical absorption measurement is not a direct measure of the absorption in the thin Si slab as the underlying Si substrate also contributes to measured absorption. Nonetheless, absorption measurements provide a useful method to characterize reflection losses and to indicate the close correspondence between experimental measurements and simulations.

The absorption enhancement of nanostructured Si films can be directly determined by analyzing spectral response photocurrent measurements (Fig. 2d-f). Note that the underlying silicon dioxide layer induces multiple Fabry-Perot resonances but do not contribute to the overall photocurrent. The photocurrent spectral response of rectangular resonators (Fig. 2e) shows a substantial overall enhancement compared to the bare film despite the sparse resonances. However for rectangular resonators, the spectral response is relatively low at the 580 nm - 670nm spectral region and the response is significantly reduced at wavelengths longer than 750 nm. By contrast, resonators with trapezoidal shape exhibit a broadband increase in photocurrent spectral response, with additional resonances observable up to 950 nm (Fig. 2f). The drop in the photocurrent density from 870 nm to 1000 nm is due to decreased solar irradiation arising from water-related absorption when the measured spectral response is weighted by the AM1.5G solar spectrum. The peak positions for simulated photocurrent spectra exhibit an excellent match to those in experimental spectra and clearly demonstrate the presence of several coupled resonances that enhance the photocurrent. The peak widths in the experimental photocurrent are found to be broader at long wavelengths compared to those in electromagnetic simulations. We attribute this discrepancy to our optical measurement system which excites the sample with illumination over the finite angular distribution emerging from a 0.14 NA focusing objective lens and also experimental deviations from the rectangular cross sections assumed in the simulations, e.g., due to slight rounding of the corners and edges of nanostructure sidewalls.

**Spectral contribution of resonant modes to absorption and carrier generation**

The integrated short circuit current density ($J_{sc}$) for the rectangular and trapezoid patterns are 7.8 mA/cm$^2$ and 12.0 mA/cm$^2$ respectively, which are considerably higher than the bare film $J_{sc}$, 3.2 mA/cm$^2$. To understand the photocurrent enhancement mechanism, we calculated spatial absorption maps at several resonant frequencies for an incident plane wave illumination polarized along the *x-axis* (Fig. 3). The maps at the same time provide the electric field distribution, $|\mathbf{E}|^2$, therefore give us an idea about the resonant modes. The spatial absorption maps are shown for different cross-sections at each wavelength, i.e. at the resonator- thin film interface, across the ridges, along the ridges, and along the ridges with 100 nm displacement from the center of the ridges. From these maps, we can identify three types of distinct resonant modes that account for photocurrent enhancement. Localized Mie resonant modes are excited by plane wave illumination at 418 nm. At an excitation wavelength of 686 nm, the dominant absorption mechanism is via waveguide modes of the trapezoid ridges propagating along the x-axis. As the Si absorption is reduced at longer wavelengths, long-range interactions between trapezoidal unit cells get stronger, and the coupling efficiency to the guided modes increases. At 948 nm illumination, waveguide modes of the thin film are excited. As seen in Fig. 2, despite efficient coupling to waveguide modes and a high (~40x) observed absorption enhancement compared to a bare Si film, the contribution of the waveguide modes to the short circuit current density is negligible due to the smaller absorption coefficient of Si near its band edge and a reduced solar Irradiance at long wavelengths. Thus in an optimization scheme for the overall performance solar cells, it is essential to consider the contribution of resonant modes to carrier generation rather than enhancement factors relative to planar thin films.

We integrated the charge carrier generation rate over all wavelengths at each point to determine the spatial distribution of the resonance contribution to photocurrent generation. Figure 4e and 4f shows the spectrum-integrated carrier generation map for a rectangular pattern. There is a pronounced pattern of modes for carrier generation, a result of the finite number of resonant modes that have symmetric profiles within the rectangular resonator pattern. We also observe a strong polarization dependence of the carrier generation in the *xz* and *yz* cross-sections. Note that although the spatial distribution of photocurrent density exhibits a local polarization dependent variation at different cross-sections, the spatially-integrated current density has no polarization dependence. Figure 4g and 4h shows the carrier generation map for the trapezoid pattern. There is a high carrier generation rate within the trapezoids while the active layer has a uniform background generation throughout its volume. The stronger contribution is attributed to the Mie resonances and ridge waveguide modes that are mainly excited within the trapezoid while uniform background contribution is due to excitation of multiple waveguide modes with spatially non-uniform mode profiles. An intriguing characteristic of the trapezoid pattern is the polarization dependence of the generation rate observed at different cross-sections. The crossed trapezoid pattern has similar carrier generation rates for both *xz* and *yz* cross-sections, which is a result of the broken symmetry that facilitates the mode hybridization and conversion between different symmetries. Films with roughened surfaces and randomly corrugated features exhibit similar light trapping effects[25]. However the trapezoidal design is distinct from non-periodic and randomly distributed structures since the non-uniformity enabling broadband absorption is implemented within a single wavelength-scale structure.

**Angular Dependence of Spectral Response**

An important parameter that determines the overall energy output of a solar cell is its angular response, since solar modules with fixed axis experience a change of sunlight incidence angle in the course of a day and throughout the year. We have measured the angular dependence of the spectral response for rectangular and trapezoid structures (Fig. 5). For both structures, the peak positions of the resonances are preserved with slight shifts at shorter wavelengths (400 nm – 650 nm) while larger shifts are observed at longer wavelengths (700 nm – 1000 nm). The observed dependence of the peak shift on wavelength is a consequence of Mie resonances mainly residing at short wavelengths with less angular dependence while waveguide modes are excited at relatively longer resonant wavelengths via Bragg-scattering which has a significant angular dependence. Figure 5c shows the overall integrated short circuit current density for each system. Trapezoid and rectangular structures show a 9% and 4% decrease in current density at an incident angle of 60º while a bare film has a 8% decrease. Optimizing structure design to give higher photocurrent density at larger incident angles can circumvent the small decrease in the photocurrent density, and thus a possible future research direction is to design structures that compensate the geometrical cosine factor to achieve a flat photocurrent density response throughout the day.

# Discussion

So far we have discussed the performance of a specific thin film absorber that incorporates trapezoidal shaped nanoresonators optimized to maximize the overall photocurrent density in 210 nm Si thin films. The proposed geometry could be optimized for a wide range of solar cell materials over a large thickness range. However the use of brute-force full-field simulations to scan a vast design parameter space would be inefficient and computationally intensive. It is preferable to use general principles that yield first order estimates for the design parameters. The modal dispersion and the periodic length scale determine the spectral position of the waveguide resonances, whereas Mie resonances can be tuned by changing the long and short cross-sections of the trapezoidal resonators. In the present fabricated design, we have chosen a period of 450 nm, providing waveguide resonance response function centered around 550 nm for 85 nm thick Si films, whereas the large variation in the trapezoid width creates Mie resonances across a wide portion of the optical spectrum, but here mainly targeted at shorter wavelengths. The design parameters can be adjusted according to the material and the thickness. For instance for a micron-scale Si thin film it would be more desirable to minimize reflection at short wavelengths and to shift the coupling resonances to longer wavelengths. AR effect could be implemented by using trapezoid structures with triangular cross-section in this case to enable an adiabatic index change and minimize the reflection loss, and the resonances could be tuned by increasing the period and the trapezoid size parameters accordingly.

In conclusion, we have proposed a semiconductor metasurface design to realize broadband absorption enhancement by efficient coupling into both broadband Mie resonators and guided mode resonances across the solar spectrum. We have fabricated a prototype nanostructured thin film cell design on silicon-on-insulator films with Schottky barrier collection of carriers, and achieving a 4-fold increase in the photocurrent compared to unetched planar thin films. The present work provides a useful framework on how to identify and analyze the resonant modes of Mie nanoresonators by generating spatial absorption maps and gives a general design strategy for use of broadband subwavelength dielectric resonators to optimize photocurrent density.

## Methods

**Sample fabrication.** Schottky barrier photodetectors as shown on Fig. 1e were realized in the 210-nm thick Si layer of a lightly doped (1 ohm cm) p-type silicon on insulator wafer using standard photolithography techniques. The electrical contacts of the detector were spaced by 120 μm. We generated arrays of Si nanoresonators between the contacts with rectangular and trapezoid shapes using standard electron-beam lithography, Reactive Ion Etching, thermal oxidation and thermal diffusion doping techniques. Optical measurements were carried out using an inverted microscope coupled with a spectrometer and an Electron Multiplication Charge Coupled Device array. Samples were illuminated using a 2X objective with a numerical aperture of 0.06 to ensure a near normal incidence. The reflected light is calibrated by a broadband dielectric mirror (Edmund Optics). Photocurrent measurements were carried out using a Fianium supercontinuum white-light source coupled to a Spectra Physics monochromator. The photocurrent signal was measured using a Keithley SourceMeter connected to a SRS lock-in amplifier. The 50x50 μm$^2$ nanoresonator arrays were illuminated with a Gaussian beam using a 5X Mitutoyo objective providing a spot size of 5 μm in diameter.

**Full-field simulations on non-periodic arrays.** Full-field electromagnetic wave calculations are performed using Lumerical, a commercially available Finite-Difference Time-Domain (FDTD) simulation software package. The FDTD simulations utilized the resonator shapes that were obtained from digitized SEM images of a 2×2 unit cell of experimentally fabricated arrays using Lumerical's digitization feature. For the material data of Si and SiO$_2$, tabulated data from Palik[26] and a constant refractive index of 1.45 were used, respectively. A mesh override region of 2×2×2 nm$^3$ was defined over the structure to amply resolve fine features in digitized SEM images.

## Acknowledgements

This work was supported by the Multidisciplinary University Research Initiative Grant (Air Force Office of Scientific Research, FA9550-12-1-0024) and used facilities supported by the DOE 'Light-Material Interactions in Energy Conversion' Energy Frontier Research Center under grant DE-SC0001293 and the Kavli Nanoscience Institute (KNI) at Caltech. We thank Dr. Dennis Callahan for helpful discussions.

## Contributions

All authors contributed to all aspects of this work.

# Figure Captions

## Figure 1

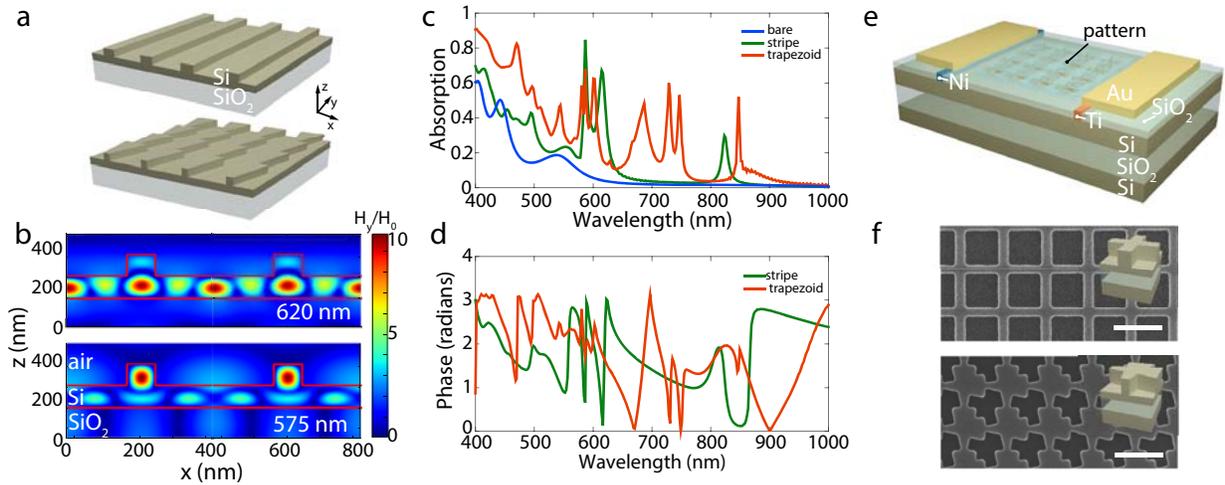

**Figure 1:** (a) A schematic representation of Si ridge grating model structure with rectangular stripe (top) and trapezoid (bottom) cross-section. (b) Simulated H-field intensity distribution maps in *xy* plane of the rectangular stripe array for plane wave incidence in TM-mode at 620 nm and 575 nm respectively. (c) Simulated absorption spectrum for bare (blue) and nanostructured Si films with rectangular stripe (green) and trapezoidal (red) ridges. (d) Calculated reflection phase from Si metasurfaces with rectangular stripe (green) and trapezoidal (red) ridges. (e) Schematic of silicon-on-insulator lateral Schottky device with embedded nanostructures. The regions where Au contact pads (yellow) are contacted with Si are indicated with blue and red colors for Ohmic and Schottky contacts, respectively. (f) SEM images of fabricated device featuring crossed rectangular (top) and trapezoidal (bottom) nanostructures. Scale bar is 0.5 μm. Insets are the corresponding unit cells of the nanostructured patterns.

**Figure 2**

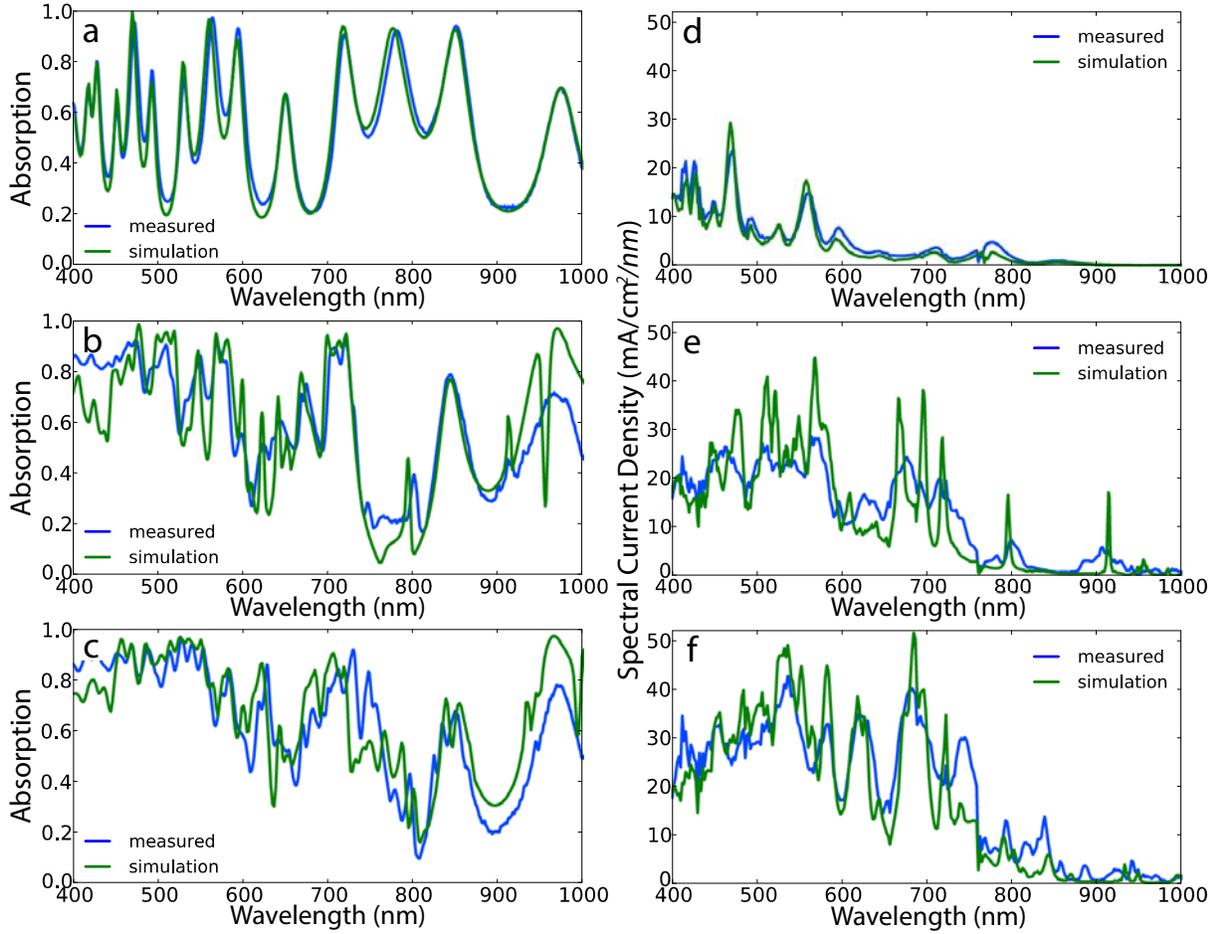

**Figure 2:** (a-c) Spectral absorption measurements and simulations (a-c); comparison of bare Si slab (a), rectangular (b) and cross-trapezoid (c) devices. Optical absorption is calculated as 1-Reflection. (d-f) Spectral photoresponse comparison of (d) bare Si slab, (e) rectangular and (f) trapezoidal devices. We assumed unity internal photocarrier collection efficiency for the simulations.

**Figure 3**

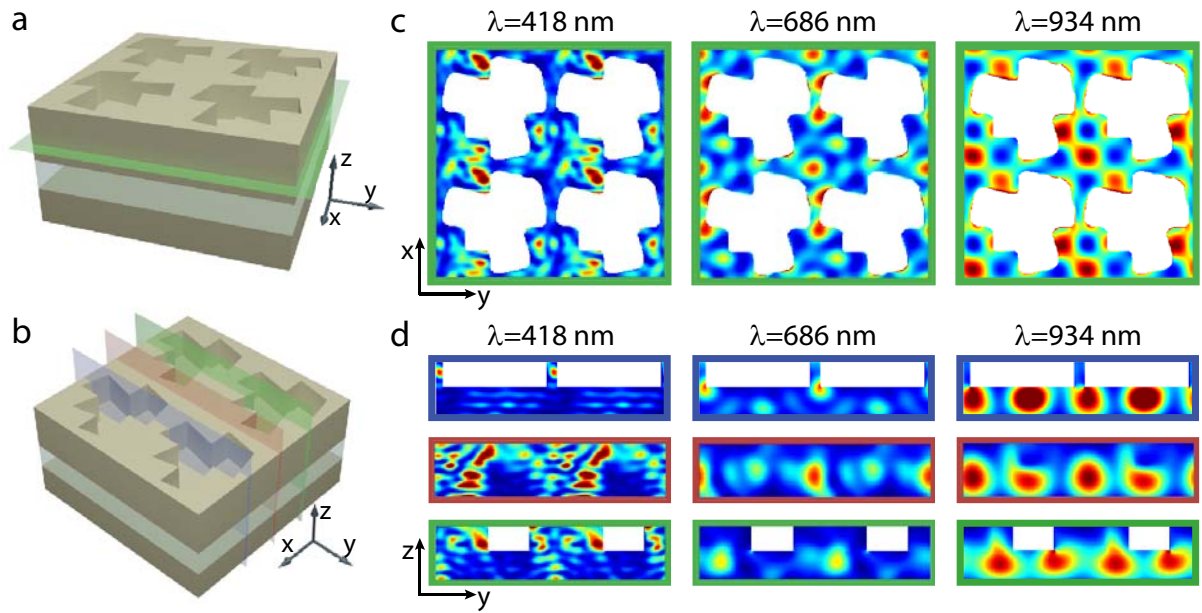

**Figure 3:** Modal analysis of the enhanced absorption. Depiction of planes (a,b) used for two-dimensional representation of modes displayed at locations on cross-sections in (c) and (d) respectively. (c) *xy* plane cross-section normalized absorption maps at various resonance wavelengths as indicated at the top. (d) *yz* plane cross-section normalized absorption maps. Map frame colors match the specified positions in (b). For each resonance, plane intersecting the narrow tip, center and wide end of the trapezoid are presented in blue, red and green frames, respectively.

**Figure 4**

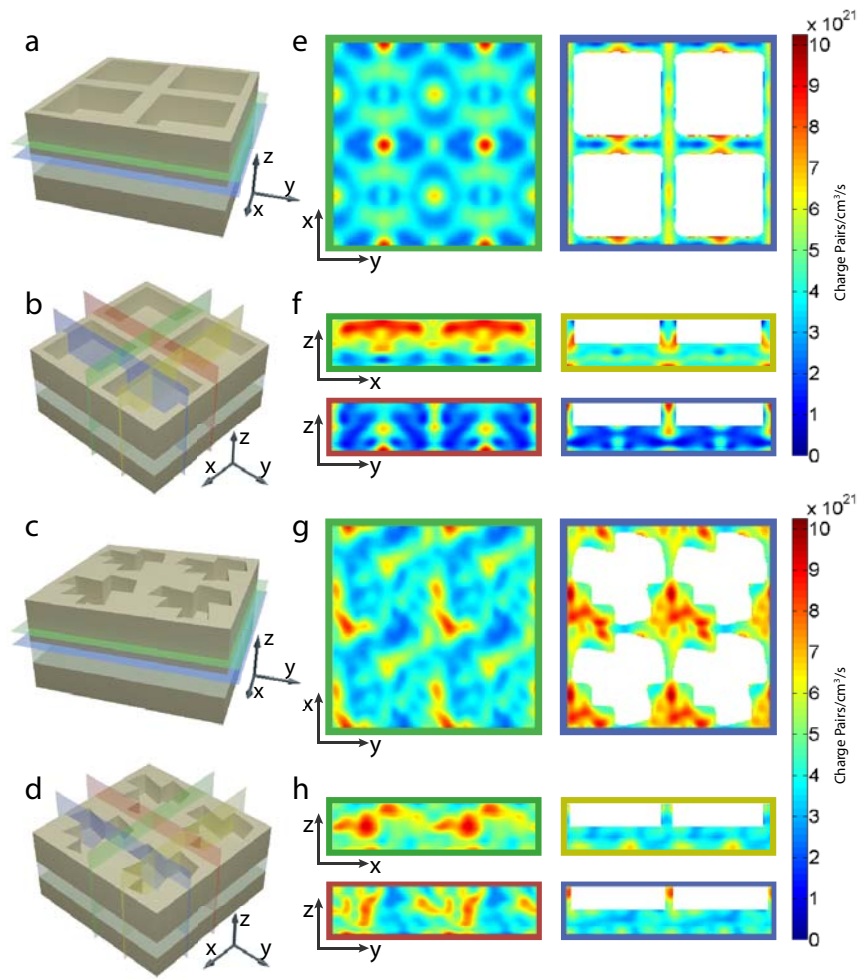

**Figure 4:** Photo-generated carrier density maps. (a-d) Depiction of locations for cross-sections for (a-b) rectangular and (c-d) trapezoidal resonator structures displayed in (e-f) and (g-h) respectively. The cross-sectional carrier generation maps (e-h) are color framed to match specified planes indicated in (a-d). Plane wave illumination was polarized along the *x* direction.

**Figure 5**

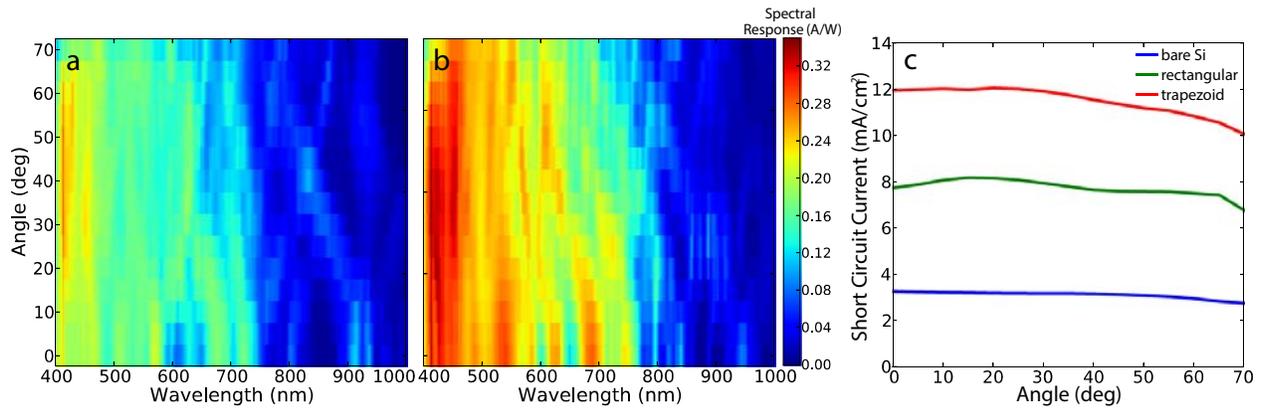

**Figure 5:** (a-b) Maps of measured spectral response from Si devices with (a) rectangular and (b) trapezoidal nanostructures versus incident angle and wavelength of the incoming light. (c) Angular dependence of Integrated $J_{SC}$ curves for bare (blue), rectangular (green), and trapezoid (red) devices.

**Contents Figure**

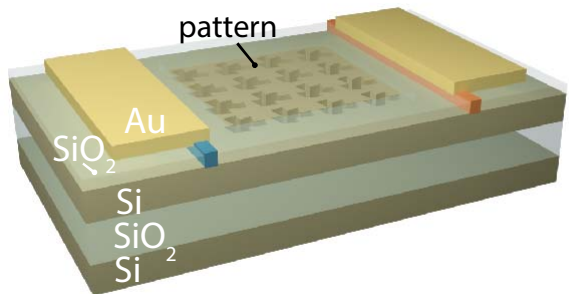
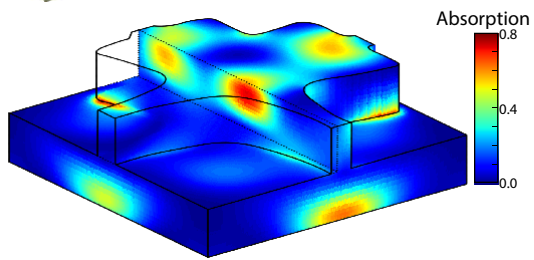
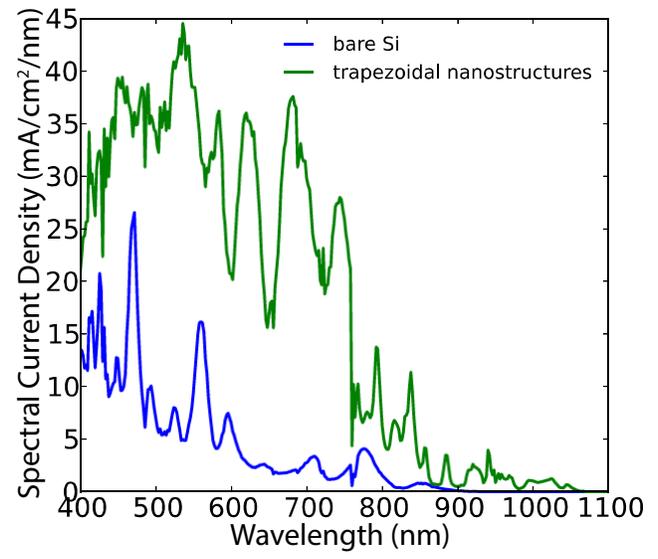